# He 3-1475 and Its Jets[a]

M. Bobrowsky[1], A.A. Zijlstra[2], E.K. Grebel[3], C.G. Tinney[4], P. te Lintel Hekkert[5], G.C. Van de Steene[6], L. Likkel[7], T.R. Bedding[8]

[1] CTA INCORPORATED, 6116 Executive Blvd., Rockville, MD 20852, USA (mattb@cta.com)

[2] ESO, Karl-Schwarzschild-Str. 2, D-85748 Garching bei München, Germany (azijlstr@eso.org)

[3] Sternwarte der Universität Bonn, Auf dem Hügel 71, D-53121 Bonn, Germany (grebel@astro.uni-bonn.de)

[4] Anglo-Australian Observatory, PO Box 296, Epping. NSW 2121, Australia (cgt@aaoepp.aao.gov.au)

[5] Parkes Observatory, Australia Telescope National Facility, P.O. Box 276, Parkes NSW 2780, Australia (plintel@atnf.csiro.au)

[6] ESO, Casilla 19001, Santiago 19, Chile (gsteene@eso.org)

[7] Washington State University, Program in Astronomy, Pullman, WA 99164-3113, USA (likkel@beta.math.wsu.edu)

[8] School of Physics, University of Sydney 2006, Australia (bedding@physics.usyd.edu.au)

# ABSTRACT

We present spectra and high-resolution images taken with HST, the NTT, the VLA, and the MPIA/ESO 2.2m of the emission-line star He 3-1475 which we suggest is a post-AGB star. The star is presumed to be at the origin of a 15″ long structure containing symmetrically opposing bright knots. The knots have radial velocities of $\pm 500$ km s$^{-1}$ from the center of He 3-1475 to the ends of the jets. HST snapshots show that the core of He 3-1475 is unipolar with a star at the SE end and the nebula fanning out toward the NW. VLA observations show the presence of OH masers, which are positioned parallel to the optical jets. A model is proposed that accounts for all of the observational data. This unusual object may link the OH/IR stars having extreme outflow velocities with highly bipolar planetary nebulae.

*Subject headings:* ISM: planetary nebulae: general — ISM: planetary nebulae: individual (He 3-1475) — stars: circumstellar matter — stars: mass loss

## 1. Introduction

He 3-1475 (=IRAS 17423−1755) is one of the few probable examples of a post-AGB star. It is an emission-line object (Henize, 1976) with unusual IRAS colours (Parthasarathy & Pottasch 1989) that place it mid-way between planetary nebulae (PNe) and HII regions in the diagram of Pottasch et al. (1988).

Riera et al. (1993) reported He 3-1475 to have high-velocity outflows and a highly dense inner region ($10^9$–$10^{15}$ cm$^{-3}$) which they interpreted as a cool circumstellar disk. He 3-1475 is probably related to a group of young PN with jet-like outflows, of which the prime example is CRL 618 (Gammie et al. 1989). Unlike either emission-line stars or PN, He 3-1475 exhibits OH emission similar to (but distinct from) OH/IR stars: the OH 1667 MHz emission profile contains features similar to the strongly bipolar high-velocity nebula OH 231.8+4.2 (te Lintel Hekkert 1991) and to three similar objects, IRAS 16342-3814, IRAS 19134+2131, and W43A (Likkel et al. 1992). The evolutionary status of these objects is still very poorly known: in particular, it is not known whether they are related to the young PN with jet-like outflows.

In this paper we present new data taken with HST, the VLA, the NTT, and the MPIA/ESO 2.2m. These confirm the outflow reported by Riera et al., and show evidence for a jet with high velocities. This is the first object showing both OH emission and an ionized outflow and may be the evolutionary link between the two groups of outflow sources.

## 2. Observations

*HST:* He 3-1475 was observed in 1993 with the Planetary Camera (PC6, image scale of 0.043″/pixel) as part of a snapshot survey of suspected proto-PNe. Two 80-s exposures were taken with each of two filters: H$\beta$ (F487N) and [O III] 5007Å (F502N). The exposures were averaged and deconvolved using 100 iterations of the Richardson–Lucy algorithm.

*VLA:* We observed both circular polarizations of the 1667MHz OH emission in 1988 November with the VLA in B-configuration. The integration time was about 10 minutes per polarization. The 63 independent channels gave velocity and angular resolutions of 1.1 km s$^{-1}$ and $1.3'' \times 1.1''$ respectively; the noise level of the total-intensity channel maps was 19 mJy/beam. A radio continuum image at 3.6 cm was made in 1993 September with the VLA in D-array. The total integration time was 10 min, giving a noise level of 0.10 mJy/beam and an angular resolution of about $10''$.

*Ground-Based Optical and near-IR Imaging:* In 1993 September we used the ESO 3.5m NTT with EMMI/RILD (image scale $0.35''$/pixel) to obtain images through Bessell I and two narrow band filters, one centered on H$\alpha$ (ESO#596, $\lambda_c = 6547.2$ Å, FWHM 73.2Å, also covering the nearby[NII] lines) and one redshifted to measure only continuum emission (ESO#598, $\lambda_c = 6665.0$Å, FWHM 6.64Å). The continuum image was subtracted from the H$\alpha$+[NII] image. Images at $K$ band were obtained with the SHARP camera (Eckart et al. 1991) on the NTT in 1994 April and at $K'$ with the COME-ON PLUS/SHARP adaptive optics camera on the 3.6m telescope in 1994 July. In both cases, the source was unresolved at $0.8''$.

*Spectroscopy:* The spectrum was obtained in 1993 September at the 2.2m MPIA/ESO telescope using EFOSC II (grism #6) and a $1''$-wide slit. The spectrum covers 4600–7200 Å at a dispersion of 2.6Å/pixel. The slit was aligned along the long axis of the object (position angle 135°).

## 3. Description and physical parameters

The NTT images (Figure 1, Plate xx) show two jet-like components extending outward from a very bright central source; the total extent is about 15″. The jets are faint in the $I$-band but bright in H$\alpha$, especially after continuum subtraction. The structure in H$\alpha$ contains three pairs of knots, each pair symmetrically opposed on either side. The pairs are at slightly different position angles, giving the impression of symmetric curvature along the jet. The emission in $I$ peaks between the H$\alpha$ knots.

The HST images are also shown in Figure 1. The jets are too faint to be seen with these exposure times and filters. However, the bright central component seen in the NTT images is resolved into a highly asymmetric structure a few arcseconds across. The off-center point-like source contains 15–20% of the total flux (about $1.1 \times 10^{-15}$ W m$^{-2}$ through the 26Å-wide [OIII] filter). It appears to show a 'plateau' of excess emission to the side. The H$\beta$ and 5007-Å images show only a few differences, such as the relative brightness of the 'plateau'. (In the H$\beta$ image there is a faint suggestion of a ring of gas viewed almost edge-on NW of the star.) The symmetry axis of the HST images is the same as the position angle of the more extended jet. This immediately rules out interaction with the interstellar medium (ISM) as the cause of the structure of the HST nebulosity.

Spectra taken along the position angle of the jets (Fig. 2) show a rich emission-line spectrum. In the center the emission is dominated by H$\alpha$ and by FeII and FeIII lines, indicating high density. In the jets the iron lines are absent, H$\alpha$ is weak and the dominant lines are 6548,6584,5755Å [NII], 6717,6731Å[SII] and 6300,6364Å [OI]. The ratio of 6584Å [NII] to H$\alpha$ is as high as 7 in the jet: the observed lines and line ratios indicate shock ionization.

A large velocity shift is apparent in the forbidden lines on either side of the center, with a total velocity width of 1000 km s$^{-1}$ (Fig. 3). The SE jet is red-shifted and the NW blue-shifted. All spectral lines show similar velocity profiles: the velocity along the jet increases linearly with distance from the center out to about 5″ and then turns over. In the inner 3″ only the hydrogen lines are detected, and they are dominated by the strong central component: there is no contribution from the jet.

The OH emission (Fig. 4) is spatially unresolved at the extreme velocities and extended (about 2″) near the central velocity (47 km s$^{-1}$). The OH emission is thus slightly more extended than the diffuse emission in the HST images but much less extended than the jets. The extreme OH emission velocities are +70.9 and +21.5 km s$^{-1}$, implying an expansion velocity of at least 25 km s$^{-1}$. This is somewhat higher than typical for AGB stars but not unusually so. The blue-shifted and red-shifted peaks are displaced by about 1″ SE–NW, indicating an aspherical shell. The direction is the same as the optical jet, but the velocity gradient is reversed. The strong peak (at 45 km s$^{-1}$) superposed on the emission profile near the central velocity (see spectrum in te Lintel Hekkert, 1991) is located in the SE direction, and appears to be spatially unresolved. The IRAS fluxes, which peak near 60$\mu$m, imply a *detached* shell with an inner radius of a few times 10$^{16}$ cm. The density of the shell would be appropriate for OH emission, and (for a distance of a few kpc), the position of the OH emission is consistent with this radius.

Interestingly, central velocities derived from the different lines do not agree — H$\alpha$, +66; H$\beta$, +143 km s$^{-1}$, where all velocities are in the LSR frame. Depending on the kinematics and geometry of the dust shell, the central velocity derived from the OH data is between 20 and 45 km s$^{-1}$ with respect to the LSR. The estimated uncertainty for the hydrogen lines is 10 km/s. For comparison, interpolating the forbidden lines at both sides of the jet indicates a systemic velocity of between 0 and 50 km/s. The difference found for the hydrogen lines is typical of scattering against a medium moving away from the emission source. Assuming that the observed profile contains both scattered and direct light, the scattering component would be larger in the blue (H$\beta$) than in the red (H$\alpha$). This is consistent with the observed velocity shift. If this explanation is correct, the implication is that *the HST images in figure 1 show mainly scattered light.* A consequence is that the emission and scattering layer move with high relative velocities — higher than the velocity of the OH-emitting shell which presumably causes the scattering. Thus, if this is true, the ionized material is moving and may be participating in the jet outflow.

The radio continuum observations found an unresolved source ($< 10''$) with a flux of 0.3 mJy, tracing free–free emission from this ionized gas. The flux density would imply an H$\beta$ flux of $2 \times 10^{-13}$ erg cm$^{-2}$ s$^{-1}$. The observed H$\beta$ flux is much higher which means that the radio continuum is optically thick at 3.6 cm. At an electron temperature of $10^4$ K this implies a diameter of the radio-emitting region of $0.03''$. Thus, the ionized material in the jet arises from a highly compact component.

The OH emission is very faint with a peak brightness temperature at the extreme velocities of $6 \times 10^4$ K. This low value suggests that the emission is stimulated but not saturated. An unsaturated maser is very sensitive to stimulation by background emission which it amplifies exponentially. The strong maser spike at the SE side is easiest to explain as background amplification, either from the central star or from the ionized component. The latter is the more likely candidate since at 1667 MHz it will be about $10^3$ times brighter than the star.

In Figure 5 we plot OH radial velocity versus projected distance. The strong trend can be explained using a model by Shu et al. (1991) for bipolar outflows in star-formation regions. Their model shows that, if a momentum-driven wind plows into an envelope in which the density decreases as r$^{-2}$, then the velocity of the envelope increases with radius and, for a well-collimated outflow, the observed velocity gradient is linear. This same model has been successfully applied to another remarkable source, presumably in transition from AGB to PN phase: HD 101584 (te Lintel Hekkert et al. 1992). An alternative but more arbitrary explanation for the velocity gradient involves a ballistic model in which 'shrapnel' is ejected from the central source with a range of expansion velocities during a single event.

The sulfur line ratios indicate densities in the jet around $10^3\,\mathrm{cm}^{-3}$, with the higher densities in regions where the intensity of the lines is lowest indicating that collisional de-excitation becomes important. The [NII] lines indicate a far higher density of $> 5 \times 10^5\,\mathrm{cm}^{-3}$ (lower densities would imply unrealistically high electron temperatures). Thus, we are led to a model where the jet contains high-density knots, separated by lower-density regions where the sulfur lines originate. The $I$-band image shows the strongest emission between the [NII] knots, and may trace the same gas as the sulfur lines. In the center the forbidden lines disappear completely, indicating that collisional de-excitation is even stronger here. The iron lines indicate densities in excess of $10^6\,\mathrm{cm}^{-3}$.

The spectrum is indicative of shock ionization, however at fairly low velocities. There is no indication for interaction with an ambient medium which would produce a much higher velocity shock. The high degree of symmetry in the jets also suggests that there is little interaction with the interstellar medium.

### 3.1. A geometric model

We find that He 3-1475 contains at least two different outflow components surrounding the central star: the fast, bipolar jet and the relatively slow detached shell traced by the OH. There is also the ionized core responsible for the radio and hydrogen-line emission. We will now combine all data sets to arrive at a source model.

The detached shell is punctured in at least two directions, namely the axis of the fast jet. We therefore propose that the shell has a torus-like structure. The HST nebulosity, being scattered light, is expected to show the innermost edge of the detached shell. The reflection is observed to be on the side where the jet comes towards us (and where the line of sight to the backside of the shell suffers the least extinction), consistent with this model. The scattering model thus explains the asymmetric appearance of the nebulosity.

As explained above, the source of the hydrogen lines (i.e. the ionized core) appears to move at high velocities with respect to the detached shell, suggesting it is associated with the jet. It is possible that we see shock ionization at the innermost (most recent) condensation in the jet.

We suggest that the brightest component in the HST images is the central star and is coincident with the center of the OH emission. In this case the brightest OH peak is offset towards the SE and may be amplifying continuum emission from the jet. The OH emission is associated with the torus, as shown by the fact that the red-shifted and blue-shifted OH velocity peaks lie in the opposite direction from the red- and blue-shifted jets. It should be noted that the positional alignment between radio and optical images is uncertain by as much as an arcsecond. A schematic diagram of He 3-1475 is shown in Figure 6.

The OH emission is unusual for an OH/IR star in that only the 1667 MHz emission is observed. This transition is favored in warm gas with short amplification path lengths (Field 1985) which, together with the faintness of the emission, implies that the OH-emitting shell is thin. A thin shell could indicate either little molecular gas (typical for post-AGB stars) or that the combination of density, temperature and molecular abundances needed for inversion of OH are only achieved at a limited range of densities. The latter situation occurs in HII regions, where the OH generally traces the surface layer of a high-density region where $H_2O$ is dissociated. Note that the highest densities found in this source ($> 10^6$ cm$^{-3}$) would lead to collisional quenching of the OH maser. The velocity structure of the OH can be explained as a slow, collimated outflow: this effect would be obtained if the OH emission occurs close to the interface of the torus and the jet rather than in the center of the torus.

If He 3-1475 is 1 kpc away, the 10″ extension of each jet implies a linear extent of $1.5 \times 10^{17}$ cm / cos $i$, where $i$ is the inclination from the plane of the sky. A velocity of 500 km s$^{-1}$ would imply an age of 95/cos $i$ yr. The angular motion could be due to precession or orbital motion. In the latter case, the origin of the jet involves a secondary star.

## 4.   Classification

The proposed model does not allow an unambiguous classification of He 3-1475. The two most likely possibilities are a post-AGB star (evolving into a PN) or a young stellar object evolving towards a naked T Tauri star. The jet-like structures are observed in both classes of objects, although He 3-1475 has peculiarities in either case.

The highly developed jets are similar to HH jets, although their velocity is very high for such a classification. The OH emission would be highly unusual for young HH-objects, as is the expanding torus. The location of He 3-1475 also does not coincide with any region of known star formation, although it is sufficiently far from the plane that no CO surveys have included this region. The variability index of the IRAS measurements (34%) is more consistent with post-AGB stars than pre-main sequence objects. For these reasons we favor a classification as an evolved object. Our result that the [NII] emission is very strong relative to H-alpha confirms a similar report by Riera et al. (1993) who suggest that the outflow is nitrogen enriched. This suggestion supports the possibility that He 3-1475 is a post-AGB star.

There are two groups of objects with similar outflows which are classified as evolved stars. The first group consists of the few known high-velocity outflow OH/IR stars, such as IRAS16342−3814 (te Lintel Hekkert et al. 1988), HD 101584 (te Lintel Hekkert et al. 1991), and OH 231.8+4.2. The second group are generally classified as young planetary nebulae and have hotter central stars; they are partly ionized (e.g., M1-16, M2-9). Whether the groups are evolutionarily connected is not known, although a possible transition object is HD 101584.

He 3-1475 is unique in showing both OH emission and an ionized outflow. It is the best case linking the two groups and suggests that objects like IRAS 16342−3814 could indeed evolve into highly bipolar young PNe. However, we must then ask why we do not see any old PNe with an extreme bipolar appearance. As the PN phase should last much longer than the post-AGB phase, this is a significant problem. A viable post-AGB hypothesis therefore requires that the star will evolve so slowly towards higher temperatures that the nebula disperses before the star enters the 'normal' PN phase (this is expected if the central star mass is very low, $< 0.55\,M_\odot$), or that the jet will cease and disperse quickly compared to both the expansion of the detached shell and the thermal time scale of the star.

Highly focussed outflows are much easier to produce in contracting than in expanding shells. This argument leads to the assumption that jets in evolved objects involve binary systems, in which the companion directs some of the outflowing material into a circumstellar (or circum-binary) disk. This disk will then be the origin of the jet. Such a model could also be applied to He 3-1475, although there is no direct evidence for a binary star. Binary interaction or even common envelope evolution can significantly accelerate the mass loss, leading to a star with below-average mass.

In conclusion, He 3-1475 is a spectacular example of mass loss leading to a high-velocity jet. It is probably a post-AGB star and may link the two OH/IR stars having extreme outflow velocities with highly bipolar PNe.

We thank Krista Rudloff and Matt McMaster for their assistance with the reduction of the HST data and John Godfrey for his help with the figures. M.B. gratefully acknowledges the hospitality of the European Southern Observatory during his visit to Garching. His observations at ESO, La Silla were supported by a grant from NASA administered by the American Astronomical Society. The H$\beta$ and [O III] images were taken with the NASA/ESA Hubble Space Telescope and obtained at the Space Telescope Science Institute which is operated by the Association of Universities for Research in Astronomy, Inc., under NASA contract NAS5-26555. Support for this work was provided by NASA through grant number GO-3603.01-91A from the Space Telescope Science Institute.

## Figure Captions

Fig. 1. NTT contour plots and HST images of He 3-1475. The contour plots show continuum-substracted H$\alpha$ and I-band images. The image scale is $0.35''$ pixel$^{-1}$. The contours are drawn at logarithmic intervals at 15 levels between 125 and 65000 ADU for the H$\alpha$ plot and at 12 levels between 7950 and 65000 ADU for the I-band plot. The central region, as well as north and south of the center, are saturated in the I-band image. The [O III] 5007 Å and H$\beta$ images of He 3-1475 were acquired with HST's Planetary Camera. These short (80 s) snapshots shows only the central, brightest region.

Fig. 2. Spectra of He 3-1475 from the center and on the red-shifted (SE) jet, $5.6''$ from the center.

Fig. 3. Velocity profiles of He 3-1475 in different lines of hydrogen, sulphur, and nitrogen at different points along the jets. (The SE jet is redshifted.) The projected outflow velocity is higher at the ends of the jet than near the center.

Fig. 4. VLA maps in different velocity channels of the OH maser emission at 1667 MHz. Each plot is labelled with the velocity in km s$^{-1}$.

Fig. 5. The velocity of each OH maser feature plotted against the radial offset from the assumed stellar position. The area of each spot is proportional to the integrated emission of the OH maser feature at each velocity.

Fig. 6. Proposed model of He 3-1475. The points labeled 45, 21, and 71 show the position of the 45 km s$^{-1}$, 21 km s$^{-1}$, and 71 km s$^{-1}$ OH maser peaks. The shading on the NW side of the torus represents reflected light from the central star, producing the asymmetry seen in the HST images.